\documentclass[referee,mathleft]{an}
\usepackage{graphicx}
\usepackage{times}
\def\down#1{\leavevmode \lower.70ex\hbox{#1}} 
\def\lesssim{\mathrel{\down{$\buildrel < \over \sim$}}}

\begin{document}

\Pagespan{789}{}
\Yearpublication{2006}%
\Yearsubmission{2005}%
\Month{11}%
\Volume{999}%
\Issue{88}%

\title{A New Numerical Method for Solving Radiation Driven Winds from Hot  
Stars}

\author{  Michel Cur\'{e} \inst{1} \fnmsep\thanks{Corresponding author:
 \email{michel.cure@uv.cl} \newline}
\and Diego F. Rial \inst{2,1}  }

\titlerunning{A New Numerical Method \dots }
\authorrunning{Cur\'{e} \& Rial}

\institute{
Departamento de F\'{\i}sica y Astronom\'{\i}a, Facultad de Ciencias, \\
Universidad de Valpara\'{\i}so, Chile.\\
\and
Departamento de Matem\'{a}ticas, Facultad de Ciencias Exactas y Naturales, \\
Universidad de Buenos Aires, Argentina.\\
}
\received{13 Mar 2006}
\accepted{ \dots  2006}
\publonline{}


\abstract{
We present a general method for solving the non--linear differential equation
of monotonically increasing steady--state radiation driven winds. We graphically 
identify all the singular points before transforming the momentum
equation to a system of differential equations with all the gradients 
explicitly give. This permits a topological classification of all 
singular points and to calculate the maximum and minimum mass--loss
of the wind. We use our method to analyse for the first time the
topology of the non--rotating frozen in ionisation m--CAK wind, with the
inclusion of the finite disk correction factor and find up to 4 singular 
points, three of the x--type 
and one attractor--type. The only singular point (and solution passing through) that satisfies 
the boundary condition at the stellar surface is the standard m--CAK singular point. 
}

\maketitle

\section{Introduction} 

Since the launch of the first satellite with a telescope on board, 
stellar winds from hot stars have been identified as a general astrophysical 
phenomenon. These winds are driven by the transfer of momentum of the 
radiation field to the gas by scattering of radiation in spectral lines. 
The theory of radiation driven stellar wind is the standard tool to describe 
the observed properties of the winds from these stars. Based on the 
Sobolev 
approxi\-mation, Castor, Abbott and Klein (1975, hereafter "CAK") developed 
an analytical hydrodynamic steady--state model which has been improved 
(or modified) later by Friend and Abbott (1986,"FA") and Pauldrach et 
al. (1986,"PPK") and general agreement with the observations could be 
obtained (for an extended review see Kudritzki and Puls, 2000, "KP").
The success to reproduce quite well the observed winds led to the development 
of a new method to deter\-mine extra--galactic distances, i.e. the wind
momentum  luminosity relation (WML,see e.g., Kudritzki, 1998, Kudritzki and
Przybilla, 2003).  
However, in the calculations involved in the WML relation, the solution of 
the improved CAK wind (hereafter m--CAK) is \textit{not} used because the 
resulting velocity fields are in disagreement with the observations (KP). 
To avoid this problem usually an {\it{ad--hoc}} $\beta$--field velocity 
profile is assumed (see KP).
The main reason of this failure of the m--CAK model is very probably 
caused by the complex structure of the non--linear transcendental equation
for the velocity gradient and its solution scheme.

Due to the non--linearity in the momentum differential equation exist many
solution branches in the integration domain. A physically reasonable solution 
describing the observed winds must start at the stellar photosphere, satisfy 
certain boundary condition, and finally reach infinity. As there is no
solution branch that covers the whole integration domain, a solution must pass
through a singular point that matches two different solution branches. 
The integration of the hydrodynamic differential equation describing
stellar winds is therefore inevitably related to the existence of singular
points and special numerical codes have been developed to identify them. In
radiation driven winds two numerical algorithms to solve the wind's
momentum equation are used.  
The first one uses a first guess of the location of the singular point (see
e.g., FA or PPK) and then applies a root--finding routine to calculate the
value of the velocity gradient. Once the velocity gradient is known, a standard routine 
(e.g. Runge--Kutta) is employee to integrate up and downstream. When the
velocity field is attained, the lower boundary condition (see below) is
calculated and the whole process is iterated until convergence. The main 
disadvantage of this method is that a initial location of the singular point must 
be known in order to integrate the equation. 
The second method (Nobili \& Turolla, 1988) is a generalisation of the Henyey
method (Henyey et al. 1964) to handle singular points. Here the nonlinear
differential equation is linearised using a finite difference scheme. The
singular point regularity condition is treated as a movable boundary condition
(see Nobili \& Turolla for details). This algorithm needs a trial velocity
profile as initial solution and uses a Newton--Raphson method for convergence
(see Krticka 2003). The advantage of the latter in comparison with the
first one, is that no special routine is needed to find either the singular
point location or the velocity gradient. However, the main disadvantage 
is that the trial function must be "close" to the solution, this is 
important especially if more than one solution exists (see Cur\'{e} 2004).

The investigation of the mathematical features of nonlinear hydrodynamical
equations of stellar winds is not new. After Parker (1960) developed the 
theory of the solar wind, Carovillano \& King (1964) studied all the possible
mathematical solutions of Parker's equa\-tion and clearly showed that no 
physical relevant solution where neglected in Parker's origi\-nal analysis, 
i.e., only one stationary--state solution is physically acceptable. 
In radiation driven stellar winds, Bjorkman (1995) performed for the
first time a detailed topological study of the original CAK non--linear 
hydrodynamic equation. 
He found a total of five singular points and showed that the only outflow 
solution that satisfy the boundary condition of zero gas pressure in infinity
is the CAK solution. 
Another step in the understanding of all possible 
solutions of the original CAK 
differential equa\-tion has been carried on by 
Cur\'{e} \& Rial (2004). 
They analysed the topology of the CAK equation with the inclusion of the
centrifugal force term due to the star's rotational speed and also took into
account the effects of changes in the ionisation of the wind (line--force
parameter $\delta$). Indeed, another singular point could be found this way,
but the corresponding topology is of the attractor--type and hence physically 
meaningless for outflowing winds. The work by Cur\'{e} \& Rial (2004) 
therefore confirmed the CAK solution being the only physical
solution satisfying the lower boundary condition (see below) and reaching 
infinity, if the stellar rotation is taken into account.

As just mentioned, the topology of the CAK equations has been analysed in
quite some detail and several authors independently confirmed the CAK solution
being the only physical solution. However, for the improved m-CAK model, 
i.e. CAK plus the finite disk correction term, such a detailed 
analysis has not yet been performed. 
In particular in view of the recent result of Cur\'{e} (2004), who indeed 
proved the existence of additional singular points (with the corresponding 
solutions passing through) in high rotational radiation
driven m-CAK winds, analysing in detail the topology of the m-CAK 
solutions appears to be highly required. 
In addition, it is crucial to understand
the solution topology of the standard 
m--CAK model if one wants to incorporate other physical processes 
into the theory. 

In this article we present a new general numerical method that
allows to solve the wind momentum equation in a simple and straightforward 
way by transforming the equations into a system of ordinary differential
equations.
A numerical (monotonicaly increasing) wind solution for this system 
can be attained using standard numerical algorithms. 
Then, we derive a simple condition that
allows to classify the topology structure of the singular points in the
integration domain and finally we apply this method
to perform for the first time a topological analysis of the 
non--rotating frozen in ionisation m--CAK model.

The structure of the article is the following. In section \ref{sec2} 
we briefly review the non--linear differential equation for the momentum in
radiation driven theory. In section \ref{sec3}, after introducing a coordinate
transformation, we present a graphical method to find the number and 
location(s) of the singular point(s). We introduce in section \ref{sec4} 
a new physical meaningless independent variable to transform the 
wind momentum equation to a
system of differential equations where all the derivatives (with respect to
this new variable) can be obtained analytically.  
In section \ref{sec5} we linearise the equations in the neighbourhood of the
singular point(s). The eigenvalues of the matrix of the
linearised system give the  
classification topology of the singular point. Section \ref{sec6} describes 
the iteration scheme to integrate the system of differential equations 
together with a lower  
boundary condition. In section (\ref{sec7}) we apply our method by analysing 
the non--rotating frozen in ionisation radiation driven wind, including the
finite disk correction factor (m--CAK model).

\section{The non--linear differential equation \label{sec2}} 
The standard stationary model for line--driven stellar winds treats an 
one--component iso\-ther\-mal radial flow, ignoring the influence of heat
conduction, viscosity and magnetic fields (see e.g., FA). 
For a star with mass $M$, radius $R_{\ast}$, effective temperature $T_{eff}$
and luminosity $L$, the momen\-tum equation including the
centrifugal force term due to star's rotation reads: 

\begin{equation} 
v\frac{dv}{dr}=-\frac{1}{\rho }\frac{{dp}}{dr}-\frac{G M (1-\Gamma )}{r^{2}}+ 
\frac{v_{\phi }^{2}(r)}{r}+g^{line}(\rho, v', n_{E})  \label{2.1} 
\end{equation} where 
$v$ is the fluid velocity, $v' = dv/dr$ is the velocity gradient, $\rho$ is the 
mass density, $p$ is the fluid pressure, $v_{\phi } = v_{rot} \, R_{\ast} /
r$, where $v_{rot}$ is the star's 
rotational speed at the equator, $\Gamma$ represents the ratio of the radiative
acceleration due to the continuous flux mean opacity, $\sigma_{e}$, 
relative to the gravitational acceleration, $\Gamma= \sigma_{e}\, L / 4 \pi c
G M$ and the last term, $g^{line}$, represents the acceleration due to the 
lines. \\

The standard parameterisation of the line--force (Abbott, 1982) is given by 
\begin{equation} 
g^{line}=\frac{C}{r^{2}}\;f_{D}(r,v,v')\;\left( r^{2}v\; v'\right) ^{\alpha
}\;\left( \frac{n_{E}}{W(r)}\right) ^{\delta }  \label{2.2} 
\end{equation}
$W(r)$ is the dilution factor and $f_{D}(r,v,v')$ is the finite disk
correction factor. The line force parameters are: $\alpha$, $\delta$ and $k$
(the latter is incorporated in the constant $C$), typical values of these 
parameter from LTE and non--LTE calculations are summarised in Lamers and 
Cassinelli (1999,"LC", chapter 8). The constant $C$ represents the eigenvalue 
of the problem (see below) and is related to the mass loss rate ($\dot{M}$)
by:   
\begin{equation} 
C=\Gamma G M k \left( \frac{4\pi }{\sigma _{E} v_{th} \dot{M}}\right) 
^{\alpha }\;\label{2.3},
\end{equation} here $v_{th}=({2k_{bol}T/m_{H}})^{1/2}$ is the hydrogen thermal
speed, $n_{E}$ is the electron number density in units of $10^{-11}cm^{-3}$
(Abbott 1982). The meaning of all other quantities is the standard one (see
e.g., LC).  

Througout this paper, we use the original m--CAK description of the 
line--force. In his topological study of the CAK model, Bjorkman (1995) used instead the absolute 
value of the velocity gradient, to allow for the possibility of non--monotonic velocity laws. 
In Cur\'{e} \& Rial (2004) and in this work we focus on physically meaningful wind
solutions, i.e. steady--state monotonically increasing outflows.

The corresponding continuity equation reads: 
\begin{equation} 
4 \pi r^{2}\rho \;v = \dot{M}  \label{2.0} 
\end{equation} 

Thus, replacing the density $\rho$ from eq.~(\ref{2.0}), 
the m--CAK mo\-men\-tum eq.~(\ref{2.1}) with the line force given 
by eq.~(\ref{2.2}) can be expressed formally as:
\begin{equation}
F(r,v,v')=0   \label{Fformal}
\end{equation} 

\section{Singular Point Location\label{sec3}}
In this section we describe a method to obtain the locations of the singular
points. After a transformation of coordinates that takes advantage of the
functional properties of the correction factor, two functions are
obtained. Plotting this functions in a 2--dimensional phase--space, after
assuming a value for the constant $C$, the intersection points of these two
curves give the exact locations of the singular points.
 
\subsection{Coordinate Transformation\label{sec-COTR}}
We define $u$, $w$, and $w'$ as follows:
\begin{equation} 
u=-R_{\ast}/r  \label{def-u} , 
\end{equation}
\begin{equation} 
w= \; v/a  \label{def-w} ,\\ 
\end{equation}
\begin{equation} 
w'=\; dw/du  ,\label{def-w'}
\end{equation} where $a$ is the isothermal sound speed. Using these new
coordinates, the mo\-men\-tum equa\-tion (\ref{Fformal}) transforms to: 
\begin{equation}
F(u, w, w') = 0  \label{Fformal2}
\end{equation}
The full equation including (\ref{Fformal2}) for the rotating m--CAK wind 
is given in Cur\'e~(2004, eq.~[7]).
\subsection{Singular Point Conditions}
As mentioned previously, there is no solution branch that covers the entire 
integration domain ($R_{\ast} \leq r \leq \infty$). A physical solution must 
cross through a singular point, that connects the two solution branches. 
The singular point condition reads:

\begin{equation}
\frac{\partial}{\partial w'} F(u,w,w') = F_{w'}(u,w,w') = 0 \label{con-sing}
\end{equation}
Hereafter we use subindex to denote partial derivatives. 
In addition to the singularity condition the regularity condition has to be
fulfilled, which is defined by:
\begin{equation}
\frac{d}{du} F(u,w,w')= F_{u}(u,w,w') + w' F_{w}(u,w,w') = 0  \label{con-regu}   
\end{equation}
\subsection{Logarithmic Coordinates}
As the velocity field grows nearly expo\-nen\-tially near the photosphere, 
we transform $w$ and $w'$ to logarithmic variables. Compared to standard methods 
this gives a better resolution close to the photosphere which is important for the 
line formation (Venero et al.~2003).

We define logarithmic variables: 
\begin{equation} 
\eta=\ln(2 w'/w) \label{def-eta} , 
\end{equation}
\begin{equation} 
\zeta=\ln(w w') \label{def-zeta} , 
\end{equation}
and obtain for the momentum equation (\ref{Fformal2}) 
\begin{equation}
F(u,\eta,\zeta) = 0 .  \label{F1}
\end{equation}
The eigenvalue $C$ is re-scaled in these coordinates, i.e., 
\begin{equation} 
\bar{C} = C\;(a^{2}R_{\ast })^{(\alpha -1)}  \label{def-Cc}
\end{equation} 
The singularity condition $w' F_{w'}(u,w,w')$ and the 
regularity condition $F_{u}(u,w,w') + w' F_{w}(u,w,w')$ transform to: 
\begin{equation}
F_{\eta}(u,\eta,\zeta)+ F_{\zeta}(u,\eta,\zeta) = 0  \label{F2} 
\end{equation} 
\begin{equation}
F_{u}(u,\eta,\zeta) + \frac{e^{\eta}}{2} \left( F_{\zeta}(u,\eta,\zeta) - F_{\eta}(u,\eta,\zeta)\right)  = 0  \label{F3}
\end{equation}

Solving equations (\ref{F1}), (\ref{F2}) and (\ref{F3}) simultaneously 
we obtain the location of the critical point $u_c$ and $\eta_c$, $\zeta_c$. In
radiation driven winds this is possible only if we know the value of the
eigenvalue $\bar{C}$, which is fixed by a boundary condition at the stellar
surface. However, assuming $\bar{C}$ as a variable, we can find the range of
values for $\bar{C}$ that allows the singular points to exist. 
Thus, we have four unknowns, $u_c$, $\eta_c$, $\zeta_c$ and $\bar{C}$ and 
only three equations
(\ref{F1},\ref{F2} and \ref{F3}) which we solve using the
implicit function theorem. 
%
\subsection{A Graphical Method for Locating the Singular Point(s)}
Fortunately, thanks to the line--acceleration term, it is possible 
to find solutions for $\zeta$ in terms of $u$, $\eta$ and $\bar{C}$, 
{\it{directly}} from the singularity condition. 
This feature represents the {\it{keystone}} of the methodology we develop 
to obtain the location of the singular points. This is also extendable for 
any additional term in the radiation driven differential equation, e.g., magnetic
fields (Friend \& McGregor, 1984) or magne\-tically cha\-nne\-led winds
(Cur\'e \& Cidale, 2002; Owocki \& ud-Doula 2004) as long as these terms do
not {\it{explicitly}} depend on the velocity gradient $w'$.

After solving $\zeta=\zeta(u,\eta,\bar{C})$ from the singularity
condition and repla\-cing it in equa\-tions (\ref{F1}) and (\ref{F3}), 
we obtain the following two functions: 
\begin{eqnarray}
R(u, \eta, \bar{C}) = 0 \label{funR} \\
H(u, \eta, \bar{C}) = 0 \label{funH}
\end{eqnarray}
where (\ref{funR}) for the rotating m--CAK wind is given in Cur\'e~(2004, eq.~[18]).

Once a value of the eigenvalue $\bar{C}$ is assumed, the intersection of the
above defined functions in the plane $u,\eta$ gives the location of the 
singular points (see below). With the value of $u_c$, we then can calculate 
$\eta_c$ and $\zeta_c$ and proceed to integrate up and downstream. 

\section{The Equivalent System of Differential Equations \label{sec4}}
In this section, we introduce a new physical meaningless independent variable
and show that the momentum equation transforms to a system of differential 
equations with all the derivatives explicitly given.


Combining the definitions of $\eta$ and $\zeta$ (equations \ref{def-eta} and
\ref{def-zeta} respectively) we obtain the following relation between its
derivatives: 
\begin{equation}
\frac{d \zeta}{du} =  e^\eta + \frac{d \eta}{du}. \label{rel-eta-zeta}
\end{equation}
Differentiating $F(u, \eta, \zeta)$, we get:
\begin{equation}
dF= F_{u}\,du + F_{\eta} \, d\eta + F_{\zeta}\,d\zeta = 0  \label{dF}
\end{equation} 
and using (\ref{rel-eta-zeta}) in (\ref{dF}) we obtain:
\begin{equation}
dF= (F_{u} - e^\eta F_{\eta})\,du  + (F_{\eta}+F_{\zeta})\,d\zeta = 0
\end{equation}
We now introduce a new physically meaningless independent
variable, $t$, defined implicitly by: 
\begin{equation}
du=(F_{\eta}+F_{\zeta})\,dt . \label{def-t}
\end{equation}

With this new independent variable, equation (\ref{F1}) is equivalent to the following system of ordinary differential equations:
\begin{equation} 
\frac{du}{dt} = X(u,\eta,\zeta) , \label{sist1} 
\end{equation}
\begin{equation}
\frac{d\eta}{dt} = Y(u,\eta,\zeta), \label{sist2}
\end{equation}
\begin{equation}
\frac{d\zeta}{dt} = Z(u,\eta,\zeta). \label{sist3}
\end{equation}
where $X$, $Y$ and $Z$ are defined as:
\begin{equation} 
X=F_{\eta} + F_{\zeta} , \label{sistX} 
\end{equation}
\begin{equation}
Y=-F_{u} - e^\eta F_{\zeta} , \label{sistY}
\end{equation}
\begin{equation}
Z=-F_{u} + e^\eta F_{\eta} . \label{sistZ}
\end{equation}

Any solution of this system of differential equations is {\it{also}} a
solution of the original momentum equation (\ref{F1}), since if any initial
condition $(u_{0},\eta_{0},\zeta_{0})$ satisfies $F(u_{0},\eta_{0},\zeta_{0})
= 0$, then any solution of (\ref{sist1}--\ref{sist3}) verifies that
$F(u(t),\eta(t),\zeta(t))=0$. 

\section{Topology of the Singular Points \label{sec5}}
In this section we describe the steps involved in the topological 
classification of any singular point.

\subsection{Linearisation}
All the critical points of the system  (\ref{sist1}-\ref{sist3}) satisfy
simultaneously $F=0$ and $X=0$, $Y=0$, $Z=0$. Because the last three equations
are non independent between them, it is sufficient to consider only two of
them.

Given $\left(u_{c}, \eta_{c}, \zeta_{c}\right)$, i.e. a singular point, we
study the behaviour of the solutions near that point by considering small
perturbations in its neighbourhood, thus:
\begin{eqnarray}
\delta\dot{u}  & =X_{u}\,\delta u + X_{\eta}\,\delta\eta+X_{\zeta}\,\delta \zeta& +\,o\left(  \delta u,\delta\eta,\delta\zeta\right) \label{lins1} \\
\delta\dot{\eta}  & =Y_{u}\,\delta u + Y_{\eta}\,\delta\eta + Y_{\zeta}\,\delta \zeta& +\,o\left(  \delta u,\delta\eta,\delta\zeta\right) \label{lins2} \\
\delta\dot{\zeta}  & =Z_{u}\,\delta u + Z_{\eta}\,\delta\eta+Z_{\zeta}\,\delta
\zeta& +\,o\left(  \delta u,\delta\eta,\delta\zeta\right) \label{lins3} 
\end{eqnarray}
>From equation (\ref{F1}), we know $\delta u$, $\delta\eta$, $\delta\zeta$ 
satisfy the following constraint:
\begin{equation}
F_{u}\,\delta u+F_{\eta}\,\delta\eta+F_{\zeta}\,\delta\zeta\;=\;o\left(  \delta
u, \delta\eta, \delta\zeta\right)  ,
\end{equation}
and therefore we can reduce this system (\ref{lins1}--\ref{lins3}) to a 2--dimensional system, i.e.:
\begin{eqnarray}
\delta\dot{u} =&\left(  X_{u}-X_{\eta} F_{u}/F_{\eta}\right)  \delta
u+\left(  X_{\zeta}-X_{\eta} F_{\zeta}/F_{\eta}\right)  \delta\zeta 
+\,o\left(\delta u,\delta\eta,\delta\zeta\right) \\
\delta\dot{\zeta} =&\left(  Z_{u}-Z_{\eta} F_{u}/F_{\eta}\right)  \delta
u+\left(  Z_{\zeta}-Z_{\eta} F_{\zeta}/F_{\eta}\right)  \delta\zeta  
+\,o\left(\delta u,\delta\eta,\delta\zeta\right)   
\end{eqnarray}

\subsection{The Matrix $B$, Eigenvalues and Eigenvectors}
Maintaining this expansion to first order in 
$\delta u$, $\delta \zeta$ and writing the last two equations in matrix 
form, we recognise the matrix $B$,
defined as: 
\begin{equation}
B=\left.  \left(
\begin{array}
[c]{cc}%
X_{u}-X_{\eta}F_{u}/F_{\eta} & X_{\zeta}-X_{\eta}F_{\zeta}/F_{\eta}\\
Z_{u}-Z_{\eta}F_{u}/F_{\eta} & Z_{\zeta}-Z_{\eta}F_{\zeta}/F_{\eta}%
\end{array}
\right)  \right|  _{\left(  u_{c},\eta_{c},\zeta_{c}\right)  } \label{def-B1}%
\end{equation}
Because $\left(u_{c}, \eta_{c}, \zeta_{c}\right)$ are defined at a singular
point, we can use $X=0$, $Y=0$ and $Z=0$ in (\ref{def-B1}),
thereby obtaining: 
\begin{equation}
B=\left.  \left(
\begin{array}
[c]{cc}%
X_{u}-e^{\eta}X_{\eta} & X_{\zeta}+X_{\eta}\\
Z_{u}-e^{\eta}Z_{\eta} & Z_{\zeta}+Z_{\eta} \label{MatrizB}
\end{array}
\right)  \right|  _{\left(  u_{c},\eta_{c},\zeta_{c}\right)  } \label{def-B}%
\end{equation}
The matrix $B$ contains the information of the structure of the topology of
the singular points. If $\det\left(B\right)<0$, the eigenvalues of $B$ are
real and have opposite signs. In this case, $\left(u_{c}, \eta_{c},
\zeta_{c}\right)$ define a saddle type singular point (Hartman--Groessman
theorem, Aman~1990), while the eigenvectors determine the tangent vectors in
the directions of stable and unstable manifolds.

\section{Method of Integration \label{sec6}}
If we start to integrate from the singular point, the values of the gradients,
$\dot{u}$, $\dot{\eta}$ and $\dot{\zeta}$ are zero, because $X=0$, $Y=0$ and
$Z=0$ at the singular point,  and therefore, we can not leave this
point. Thus, we need to move in the neighbourhood of this singular point in the
direction of the eigenvalues and from this new position start to integrate.
\subsection{The stable manifold}
A physically stationary solution of the wind model must have a monotonically
increasing velocity profile, this is topologically characterised by the
condition $\dot{\zeta}>0$, thus we need to move in the direction of the
eigenvector of $B$ that satisfies this criterion.
Defining the components of this eigenvector of $B$ as $\left(\delta u,
\delta\zeta\right)$, we find that $\delta\eta$ fulfils:
\begin{equation}
\delta \eta=-\left( F_{u}\,\delta u + F_{\zeta}\,\delta \zeta\right)
\,/\,F_{\eta}.
\end{equation}

Once we know the eigenvector in phase--space $(u. \eta, \zeta)$, we  
integrate up and downstream starting from the points $q^{\pm}$ defined by:
\begin{equation}
q^{\pm}=\left(  u_{c}\,\eta_{c}, \zeta_{c}\right)\,\pm\, \varepsilon \,
\left(\delta u, \delta \eta, \delta\zeta\right) 
\end{equation}
where $0< \varepsilon \ll 1$. The solution that starts at the initial point
$q^{-}$, at $t=0$, achieves the photosphere, $u=-1$ at $t=T^{-}$ and the
solution that starts at the initial point $q^{+}$, also at $t=0$, reaches
infinity, $u=0$ at $t=T^{+}$.

\subsection{Lower Boundary Condition}
The location of the singular point is determined by the lower boundary
condition. In radiation driven winds from hot stars, the most frequently 
used boundary condition is a constraint on the optical depth integral: 
\begin{equation}
\int_{R_{\ast}}^{\infty} \,\sigma_{E} \,\rho(r) \,dr = \frac{2}{3} \label{tau23}
\end{equation}
An equivalent lower boundary condition is to a given value of the density 
at the stellar surface, i.e.
\begin{equation}
\rho(R_{\ast})= \rho_{\ast}
\end{equation}

Usually, in case the optical depth integral is used, it can be calculated 
only after the numerical integration has been carried on. We now show a 
calculation that permits to incorporate the boundary condition as a fourth 
differential equation of the system given by equations 
(\ref{sist1}-\ref{sist3}).  

The boundary condition at the stellar surface is given by
\begin{equation}
\gamma_{0}=\int_{-1}^{0}\phi\left(  x,\eta,\zeta\right)  dx \label{b-c}%
\end{equation}
with $\gamma_{0}$ being a fixed value. Using equation (\ref{sist1}), we can write $\gamma_{0}$ as $\gamma_{0}=\gamma_{0}^{+}-\gamma_{0}^{-}$ where
\begin{equation}
\gamma_{0}^{+}=\int_{0}^{T^{+}}\phi\left(  u,\eta,\zeta\right)  X\left(
u,\eta,\zeta\right)  dt ,%
\end{equation} 
and
\begin{equation}
\gamma_{0}^{-}=\int_{0}^{T^{-}}\phi\left(  u,\eta,\zeta\right)  X\left(
u,\eta,\zeta\right)  dt
\end{equation}

Then, we have $\gamma_{0}^{\pm}=\gamma\left(  T^{\pm}\right)$ where
$\gamma\left(t\right)$ satisfies the following differential equation: 
\begin{equation}
\dot{\gamma}=\phi\left(  x,\eta,\zeta\right)  X\left(  x,\eta,\zeta\right). \label{gpto}%
\end{equation}

Thus, adding this last equation to our system of differential equations
(\ref{sist1}-\ref{sist3}), we cal\-cu\-la\-te si\-mul\-ta\-neo\-sly the value
of the lower boundary condition while the numerical integration is performed.

\subsection{The Iteration Cycle \label{sec-itcy}}
The iteration cycle for finding out the numerical solution of (\ref{F1})
and satisfying the 
boundary condition (\ref{b-c}) is given by:
\begin{itemize}
\item Guess a value of the Eigenvalue $\bar{C}$

\item Calculate the values of $\left(  u_{c},\eta_{c},\zeta_{c},\gamma\right)
$ by solving the system (\ref{sist1}-\ref{sist3}) and (\ref{gpto}), and start
to integrate numerically from $\left(  q^{\pm},0\right)$. 

\item After the integration, compare the value of $\gamma_{0}$  with
$\gamma\left(  T^{+}\right) -\gamma\left(  T^{-}\right)  $ and modify
$\bar{C}$. 

\item Repeat the cycle until convergence is reached.
\end{itemize}

\section{Topology of non--rotating frozen in ionisation m--CAK wind \label{sec7}}
In this section we study the topology of the frozen in ionisation
non--rotating m--CAK wind, using the methodology developed in previous
sections. We solve equation~(\ref{2.1}) with the line force term given by
eq.~(\ref{2.2}) with $\delta=0$. 

In order to compare our results with previous topological analysis
(Bjorkman~1995, Cur\'e \& Rial~2004) we adopt the same B2 V star's parameters
and line--force parameters used in these studies. Table \ref{tabla1} and
\ref{tabla2} summarise these parameters. 

\begin{table}
\caption[]{B2 V Stellar Parameters} \label{tabla1}
\begin{center}
\[
\begin{array}{cccc}   
\hline \hline
\noalign{\smallskip}
R/R_{\sun} \; & M/M_{\sun} \; & L/L_{\sun}  \; & T_{eff}/K\; \\
\noalign{\smallskip}
\hline
\noalign{\smallskip}
4.5\;&9.0\;&3553.\;&21000.\;\\
\noalign{\smallskip}
\hline
\end{array}
\]
\end{center}
\end{table} 
\begin{table}
\caption{Line--force Parameters} \label{tabla2}
\begin{center}
\[
\begin{array}{ccc}   
\hline \hline
\noalign{\smallskip}
k\;& \alpha \; & \delta \\%
\noalign{\smallskip}
\hline
\noalign{\smallskip}
0.212\; & 0.56\; & 0.0\\
\noalign{\smallskip}
\hline
\end{array}
\]
\end{center}
\end{table}

\subsection{Graphical method for the location of critical points
\label{met-graf}}
The first step to understand the topology of this non--linear differential
equation is to know the location of the singular point(s). This is done using
the functions $R$ and $H$, which are given in $u, \eta, \zeta$ coordinates by:

\begin{eqnarray}
R(u,\eta) =&\Lambda^{-1} \left[ e^{\eta }\,\left( \frac{e^{2\,\eta }}{2} - \frac{2}{u^2} \right) \,
   \left( -1 + \alpha  \right) \,f_{D}(\lambda)  
+ \left( 2\,A + 8\,e^{\eta } + \frac{8}{u} +  
4\,A\,e^{\eta }\,u - e^{2\,\eta }\,u \right) \,f_{D}'(\lambda)\right] \nonumber \\ \label{fun-R}
\end{eqnarray} and
\begin{eqnarray}
H(u,\eta, \bar{C} ) =& - \, \bar{C} \,+\, \Lambda^{-1} \, e^{\eta }\,\left( A + \frac{2}{u} \right)  \, \,\left[\frac{1}{2}\,e^{\eta } + \Lambda^{-1} (A + \frac{2}{u})\, 
(\alpha \,e^{\eta }\,f_{D}(\lambda) -  2\,u\,f_{D}'(\lambda))\right]^{-\delta} \nonumber \\ \label{fun-H}
\end{eqnarray} where $\Lambda$ is an auxiliary function defined as:

\begin{eqnarray}
\Lambda= \left( 1 - \alpha  \right) \,e^{\eta }\,f_{D}(\lambda) + 
  2\,u\,f_{D}'(\lambda) \label{fun-Lambda}
\end{eqnarray} and
\begin{equation}
\lambda=u\,(u+2\,e^{-\eta }) \label{def-lambda} 
\end{equation}
The analytic expression for the finite--disk correction factor is given by:
\begin{equation} 
f_{D}(\lambda)=\frac{1}{(1-\alpha)}\frac{1}{\lambda} \left[1-\left(1-\lambda
\right)^{(1+\alpha)} \right].  \label{def-cf} 
\end{equation} 
and $f_{D}'(\lambda)$ denotes the total derivative of $f_{D}(\lambda)$  with
respect to $\lambda$.
\begin{figure}[htbp] 
\begin{center}
\includegraphics[width=0.6\textwidth]{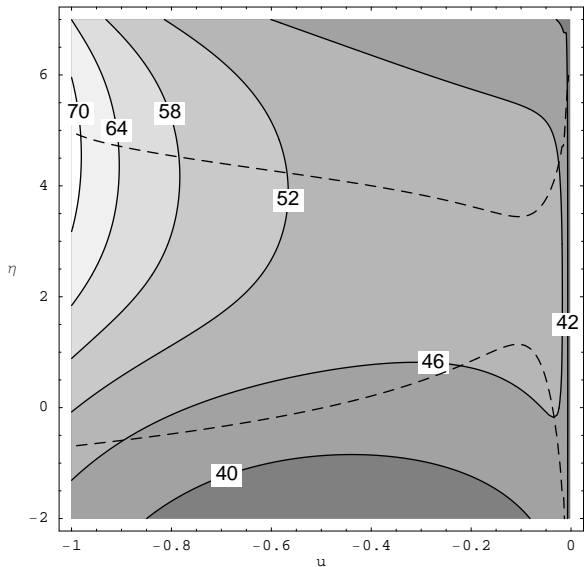}
\end{center}
\caption{The iso--contours $R(u,\eta)=0$ (dashed--line) and iso--contours
$H(u,\eta, \bar{C})=0$ (solid--lines) for different labelled values of
$\bar{C}$ as function of the variables $u$ and $\eta$. See text for
details.
\label{fig1}}
\end{figure} 

Fig.\,\ref{fig1} shows the curves defined by $R(u,\eta)=0$ (dashed lines) and
$H(u,\eta,\bar{C})=0$ (solid lines), for different values of $\bar{C}$.
The intersection of this two curves gives the location of the singular points
and the number of singular points depends on the assumed value of the
eigenvalue. We see clearly that there are {\it{two}} families of curves for
$R(u,\eta)=0$, which are independent of the assumed value of $\bar{C}$.

The standard critical point of the m--CAK model is located close to the
stellar surface, $u \lesssim -1$. On the other hand, the velocity is small
near the photosphere, therefore $\eta\,>\,0$ and the family curve of the
standard m--CAK model is the upper $R(u,\eta)=0$ curve. The value of $\bar{C}$
for the singular point located in the upper $R=0$ curve ($u \lesssim -1$,
$\eta>0$) is in agreement with a typical value of the mass loss rate of a B2 V
star, $\dot{M} \sim 10^{-9}\,M_{\sun} \,yr^{-1}$ which corresponds
approximately to $\bar{C} \sim 70$. The values of the coordinates $u$, $\eta$
and $\zeta$ at this critical point is given in table \ref{tabla3}, labelled by
$A$. \\
The more complex case of $\bar{C}=46$ ($\dot{M} \sim 2 \cdot 10^{-9}\,M_{\sun}
\,yr^{-1}$) shows that the curves $R=0$ and $H=0$ intersects in {\it{four}}
different locations, one in the upper $R=0$ curve (labelled by E) and three in
the lower $R=0$ curve (labelled from B to D). Table \ref{tabla3} also
summarises the coordinates of these critical points.
\begin{table}
\caption{Singular Points Coordinates} \label{tabla3}
\begin{center}
\[
\begin{array}{ccccccc}   
\hline \hline
\noalign{\smallskip}
Label & \bar{C}\;& r_c\,& u_c\; & \eta_c \,& \zeta_c \,\\
\noalign{\smallskip}
\hline
\noalign{\smallskip}
A\; & \;70\; & 1.018\; & -0.982\; & +4.905\; & +7.468\; \\
B\; & \;46\; & 1.116\; & -0.896\; & -0.587\; & +5.638\; \\
C\; & \;46\; & 4.505\; & -0.222\; & +0.753\; & +7.253\; \\
D\; & \;46\; & 29.41\; & -0.034\; & -0.176\; & +7.344\; \\
E\; & \,46\; & 40.32\; & -0.020\; & +4.605\; & +7.439\; \\
\noalign{\smallskip}
\hline
\end{array}
\]
\end{center}
\end{table}

An additional benefit of this graphical method is that this curves intersect
only for a range of values of the eigenvalue $\bar{C}$, therefore this method
gives a lower ($\dot{M}_{min}$) and an upper ($\dot{M}_{max}$) limit of the
mass loss rate of the wind. In this case (see Figure \ref{fig1}) it is
approximately $42 \lesssim \bar{C} \lesssim 72$, corresponding to $8.8 \cdot
10^{-10} \,M_{\sun}\,yr^{-1} \lesssim \dot{M} \lesssim 2.3 \cdot 10^{-9}
\,M_{\sun}\,yr^{-1} $. 

\subsection{The system of differential equations}
Once the location of a critical point is known we can integrate up and
downstream from this point. As we stated, the non--linear differential
equation (\ref{Fformal2}) can be transformed to the set of equations
(\ref{sistX} - \ref{sistZ}). In our case the functions $X$, $Y$ and $Z$ are: 

\begin{eqnarray}
 X(u,\eta ,\zeta, \bar{C}) =&-\,\bar{C}\,e^{\alpha \,\zeta }\,(\alpha\,f_{D}(\lambda)
                             +\,2\,u\,e^{- \eta }\,f_{D}'(\lambda) ) 
                             +\,e^{\zeta } - \frac{1}{2}\,e^{\eta } \label{eqX}
\end{eqnarray}
\begin{eqnarray}
Y(u,\eta ,\zeta, \bar{C}) =& \bar{C}\,e^{\alpha \,\zeta}\,(\alpha \,e^{\eta}\,f_{D}(\lambda)
                             +\,2\,\left( e^{-\eta}+u\right)\,f_{D}'(\lambda)) 
                            -\,e^{\zeta  + \eta } + \frac{2}{u^2} \label{eqY} 
\end{eqnarray}
\begin{eqnarray}
Z(u,\eta ,\zeta, \bar{C}) =& 2\,\bar{C}\,e^{\alpha \,\zeta }\,\left( e^{-\eta }                                        + 2\,u \right) \,f_{D}'(\lambda )  
                            -\frac{1}{2}\,e^{2\,\eta } + \frac{2}{u^2}  \label{eqZ}
\end{eqnarray}
Before we start to integrate, we determine the classification topology of 
each one of the singular points of Table \ref{tabla3}. 

\subsection{Topology classification}
The Hartman--Groessman theorem~(Aman~1990), states that a singular point
topology can be obtained from the multiplication of all eigenvalues of the
matrix $B$ or equivalently its determinant. This matrix $B$ is defined by
eq. (\ref{MatrizB}). In our case of study, $B$ is: 
\begin{equation}
B(u, \eta)= (u \Lambda)^{-1} \,\left[ f_D(\lambda )\,B_0(u,\eta ) + f_D'(\lambda )\,B_1(u,\eta ) + f_D''(\lambda )\,B_2(u,\eta ) \right]
\end{equation}
Where the matrix $B_0$ is defined as:
\begin{equation}
B_0= e^{\eta }\,u\,\left( 1 - \alpha  \right)\,\left(
\begin{array}
[c]{cc}%
\frac{1}{2}\,e^{2\,\eta }  & \left( A + 2/u \right) \,\alpha \\
e^{3\,\eta } - 4/u^3 & -e^{2\,\eta } 
\end{array}
\right)
\end{equation}
Matrix $B_1$ is:
\begin{equation}
B_1= \left(
\begin{array}
[c]{cc}%
u\,\left( e^{2\,\eta }\,u + 2\,\left( A + \frac{2}{u} \right) \,
     \Pi  \right) &
     4\,u\,\left( 2 + A\,u \right) \,\left( -1 + \alpha  \right) \\
4 - \frac{8}{u} + 2\,e^{3\,\eta }\,u^2 + 6\,e^{\eta }\,\left( 2 + A\,u \right) & 
-2\,u\,\left( e^{2\,\eta }\,u - \left( A + \frac{2}{u} \right) \,\Pi \right)
\end{array}
\right)
\end{equation}
where the auxiliary function $\Pi$ is defined as:
\begin{equation}
\Pi=\left( 1 - \alpha  + e^{\eta }\,\left( u - 2\,u\,\alpha  \right)  \right)
\end{equation}
and the matrix $B_2$ is:
\begin{equation}
B_2= \frac{4\,\left( 2 + A\,u \right) }{e^{\eta }}\,\left(
\begin{array}
[c]{cc}%
u\,\left( 1 + 2\,e^{\eta }\,u \right)  & -u^2\, \\
\left( 1 + 2\,e^{\eta }\,u \right)^2 &- u\,\left( 1 + 2\,e^{\eta }\,u \right)  
\end{array}
\right)
\end{equation}

\begin{table}
\caption{Singular points Topology classification} \label{tabla4}
\begin{center}
\[
\begin{array}{lcc}   
\hline \hline
\noalign{\smallskip}
& det(B)\;& Topology \\
\noalign{\smallskip}
\hline
\noalign{\smallskip}
A\; & \;-\; & X$--$type   \\
B\; & \;-\; & X$--$type   \\
C\; & \;+\; & Spiral      \\
D\; & \;-\; & X$--$type   \\
E\; & \;-\; & X$--$type   \\
\noalign{\smallskip}
\hline
\end{array}
\]
\end{center}
\end{table}

Table \ref{tabla4} summarises the value of the determinant of the B--matrix
for all the singular points of Table \ref{tabla3}. As expected the standard
m--CAK singular point (A) is an X--type one (or saddle--type: eigenvalues are
real, one negative and one positive). The other singular point (E) of the
upper $R=0$ family is also an X--type singular point. The lower $R=0$ family
of singular points, has two X--type singular points (B, D) and one spiral (C,
spiral--unstable: complex conjugate and  positive real part eigenvalues). 

Once the topology of each singular point is known, we integrate the 
system of di\-ffe\-ren\-tial equa\-tions in order to search for physical 
solutions of radiation driven winds. 

\subsection{Numerical Integration}
Now we show the results of the numerical integration of the system of 
differential equations given by (\ref{sist1} -- \ref{sist3}) together 
with (\ref{eqX}--\ref{eqZ}).

\subsubsection{The Upper $R=0$ Family}
\begin{figure}[htbp] 
\begin{center}
\includegraphics[width=0.45\textwidth]{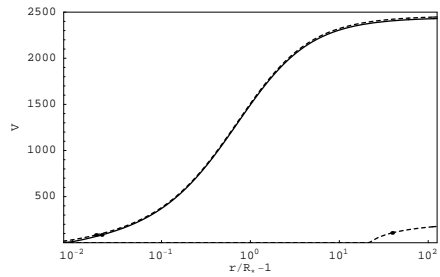}
\end{center}
\caption{Velocity (in $km \, s^{-1}$) versus $(r/R_{\ast}-1)$. The unique
curve that starts at the stellar surface and reaches infinity is the m--CAK
original solution (solid line). See text for details. \label{fig2}} 
\end{figure}

Fig.\, \ref{fig2} shows the velocity profile as function of $\log(r/R_{ast} -
1)$. 
The first iteration starting from the singular points A and D are shown in
dashed--lines.
After performing the iteration described in section (\ref{sec-itcy}), with the
lower boundary condition given by eq. (\ref{tau23}), both so\-lu\-tions
converges after few iterations to the {\it{same solution}}, shown in Figure
\ref{fig2} (solid line). This solution corresponds to the standard
m--CAK solution (FA or PPK), and the values of the location of the singular
point, eigenvalue, mass loss rate and terminal velocity are respectively:
$r_c=1.021\,R_{\ast}$ ($u_c=-0.979$), $\bar{C}=69.758$, $ \dot{M}= 9.39 \sim
10^{-10}\, M_{\sun}\,yr^{-1} $, $v_{\infty}=2441\,s^{-1}$. 

\subsubsection{The Lower $R=0$ Family}
\begin{figure}[htbp] 
\begin{center}
\includegraphics[width=0.45\textwidth]{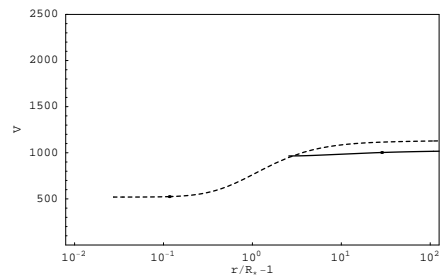}
\end{center}
\caption{Velocity (in $km \, s^{-1}$) versus $(r/R_{\ast}-1)$. The solution
that passes through singular point B (dashed--line) reaches the photosphere with 
a too high value of the velocity. While the solution passing through the singular 
point D (solid--line) does not reach the photosphere. See text for details. \label{fig3}} 
\end{figure}

Fig. \ref{fig3} show the Velocity profile of the first iteration starting
from the singular points B (dashed--line) and D (solid--line). Both
solutions reach the neighbourhood of stellar surface significantly exceeding 
reasonable values of the velocity, i.e., $V(R_{\ast}) \sim 520\,km \,s^{-1}$ and 
$V(R_{\ast}) \sim 960 \,km \,s^{-1}$ when the iteration started at point B and
D respectively.\\

The iteration algorithm described in section \ref{sec-itcy} does not converge,
because the location of the singular point is shifted beyond the integration
domain, $u>0$, for both singular starting points B and D. We therefore
conclude that the solutions starting from singular point B or D {\it{do
not satisfy}} the lower boundary condition given by eq. (\ref{tau23}).

\section{Conclusions \label{sec-CONC}}
We have developed a new method for solving monotonically 
increasing steady--state radiation driven winds which 
can be summarised as follows: i) find the location of critical points,
ii) classify its topology, iii) transform the non--linear differential
equations to a system of ordinary differential equations with all the 
derivatives explicitly given, and finally iv) iterate numerically the system 
to obtain a wind solution. 

We applied this method to study for the first time the topology of the 
non--rotating, frozen in ionisation m--CAK wind, i.e. the CAK wind 
with the finite disk correction factor.
In his CAK topological analysis, Bjorkman's (1995)
found for monotonically increasing outflow solutions only one solution
branch with one singular point and one physical wind solution, i.e. the original CAK solution.
In Cur\'{e} \& Rial (2004) we identified two solution branches with two singular points 
and one physical wind solution.\\
In this work we find two solution branches and up to four singular points, 
de\-pen\-ding of the value of the Eigenvalue. The standard m--CAK singular point, 
however, is the only one that satisfies the boundary condition at the stellar surface. 
Furthermore, in the range of possible Eigenvalues, the physical solution corresponds 
to a value close to the maximum and therefore to a {\it{minimum}} mass-loss rate.\\
We conclude therefore, that for the boundary conditions considered here, the wind
adopts the minimum possible mass-loss rate.

Our new method can be easily applied to more general cases, e.g., rotating
radiation driven winds or winds with magnetic fields. These topics are beyond
the scope of this article and we leave them for future studies.

\begin{acknowledgements}
We want to thank M.R. Schreiber for his valuable comments and help in 
improving this article. This work has been possible thanks to a research cooperation agreement
UBA\--\-UV and DIUV project 03/2005. 
\end{acknowledgements}

\end{document}